\documentstyle[prd,aps,preprint,epsfig]{revtex}

\tighten

\begin{document}
\draft
 
\pagestyle{empty}

\preprint{
\noindent
\hfill
\begin{minipage}[t]{3in}
\begin{flushright}
LBNL--48840 \\
UCB--PTH--01/37 \\
hep-ph/0106354 \\
August 2001 (revised)
\end{flushright}
\end{minipage}
}

\title{Final-state interactions and $s$-quark helicity conservation in 
$B\to J/\psi K^*$}

\author{
Mahiko Suzuki
}
\address{
Department of Physics and Lawrence Berkeley National Laboratory\\
University of California, Berkeley, California 94720
}


\maketitle

\begin{abstract}

The latest BaBar measurement has confirmed substantial strong phases for
the $B\to J/\psi K^*$ decay amplitudes implying violation of factorization
in this decay mode. In the absence of polarization measurement of a lepton
pair from $J/\psi$, however, the relative phases of the spin amplitudes
still have a twofold ambiguity. In one set of the allowed phases the 
$s$-quark helicity conserves approximately despite final-state interactions.
In the other set, the $s$-quark helicity is badly violated by long-distance
interactions. We cannot rule out the latter since validity of 
perturbative QCD is questionable for this decay. We examine
the large final-state interactions with a statistical model. Toward 
resolution of the ambiguity without lepton polarization measurement, 
we discuss relevance of other $B\to 1^-1^-$ decay modes that involve the 
same feature. 

\end{abstract}
\pagestyle{plain}
\narrowtext

\setcounter{footnote}{0}

\section{Introduction}
The BaBar Collaboration\cite{BaBar} has shown in line with 
CDF\cite{CDF} that substantial strong phases are generated in the decay 
$B\to J/\psi K^*$. It is not surprising since the argument of short-distance
dominance does not hold for this decay according to perturbative QCD 
study\cite{B,L} of final-state interactions (FSI).  Beneke et al\cite{B}
question short-distance dominance on the basis of the size of $J/\psi$, while
Cheng and Yang\cite{C} actually find a large correction to factorization 
from a higher twist in the case of $B\to J/\psi K$. 

Since the experiment does not measure polarization 
of the lepton pair from $J/\psi$, there is a twofold ambiguity left in 
the relative strong phases of three spin amplitudes. Specifically, 
the relative phase between two transverse spin amplitudes is determined 
only up to $\pi$. Two allowed set of phases are physically inequivalent 
and correspond to very different physics for FSI.

The decay $B\to J/\psi K^*$ occurs dominantly by the quark process 
$\overline{b}\to \overline{c_L}c_L\overline{s_L}$ through the tree 
decay operators. In the perturbative picture, $\overline{s_L}$ would pick 
a $u/d$-quark to form the final $K^*$. If $\overline{s_L}$ maintains its 
helicity, $K^*$ cannot be in helicity $-1$. Consequently we expect naively 
that the helicity +1 amplitudes should dominate over the helicity $-1$ 
amplitude. The twofold ambiguity left in the analysis \cite{BaBar,CDF,CLEO}
corresponds to dominance of helicity $+1$ or $-1$. If helicity $+1$ 
dominates, factorization may still be a decent approximation apart from the
strong phases. But if helicity $-1$ dominates, long-distance FSI are 
large and flip the $s$-quark helicity. Therefore it is 
important to resolve this ambiguity in order to test robustness of 
factorization and to understand the nature of FSI in general. 

When FSI is large, we have no reliable way to compute individual strong 
phases. A statistical model\cite{SW} was developed to fill the void. In this 
model large phases and helicity violation can occur if color suppression 
is severe and rescattering is strong enough in $B\to J/\psi K^*$. Guided 
by the statistical model, we look for the decay modes which share 
the same feature. Aside from $B_s\to J/\psi\phi$, we propose measurement 
of $B\to \psi(2s)K^*$, $B^-\to D^{0*}\rho^-$, and other $B\to 1^-1^-$ modes.
Though final resolution of the ambiguity requires lepton polarization 
measurement in some future, measurement of the spin amplitudes of 
these decay will help us to understand the FSI better.  

\section{Experiment}
  Three spin amplitudes $A_{\parallel,\perp, 0}$ of $B\to J/\psi K^*$ are
related to helicity amplitudes $H_{\pm 1,0}$ by\cite{spin-amp,T'Jampens}:
\begin{equation}
    A_{\parallel} = (H_{+1}+H_{-1})/\sqrt{2}, \;\;
    A_{\perp} =     (H_{+1}-H_{-1})/\sqrt{2}, \;\;
    A_{0} = H_{0}, \label{A}
\end{equation}
where helicity amplitudes are defined in the rest frame of $B$ by
\begin{equation}
 H_{\lambda}=\langle  J/\psi(\lambda),K^*(\lambda)|H|B\rangle .
\end{equation}
We follow the original sign convention of Dighe et al\cite{spin-amp}.

 Relative magnitudes of $A_{\parallel,\perp,0}$ for 
$B(q\overline{b})\to J/\psi K^*$ are given by BaBar\cite{BaBar} as
\begin{eqnarray}
 |A_0|^2 & = & 0.597 \pm 0.028 \pm 0.024 \nonumber \\
 |A_{\perp}|^2 & = & 0.160 \pm 0.032 \pm 0.014 \nonumber \\
 |A_{\parallel}|^2 &=& 1 - |A_0|^2 - |A_{\perp}|^2. \label{mag}
\end{eqnarray}
The phases are quoted in radians as
\begin{eqnarray}
 \phi_{\perp} &\equiv& \arg(A_{\perp}A^*_0)= -0.17 \pm 0.16 \pm 0.07, 
                  \nonumber \\ 
 \phi_{\parallel} &\equiv& \arg(A_{\parallel}A^*_0)= 2.50 \pm 0.20 \pm 0.08. 
                  \;\;[{\rm Solution \;I}] \label{phase1} 
\end{eqnarray}
However, since measurement of the interference terms in the angular 
distribution is limited to Re$(A_{\parallel}A_0^*)$, Im$(A_{\perp}A_0^*)$, 
and Im$(A_{\perp}A_{\parallel}^*)$, there exists an ambiguity 
of\cite{Chiang,Bernard}
\begin{eqnarray}
      \phi_{\parallel} &\leftrightarrow & - \phi_{\parallel} \nonumber \\
      \phi_{\perp} &\leftrightarrow & \pi - \phi_{\perp} \nonumber  \\
      \phi_{\perp}-\phi_{\parallel} & \leftrightarrow &
               \pi - (\phi_{\perp}-\phi_{\parallel}). 
\end{eqnarray}
Therefore, another set of values,
\begin{eqnarray}
      \phi_{\perp} & = & \arg(A_{\perp}A^*_0)= -2.97 \pm 0.16 \pm 0.07,
                  \nonumber \\
 \phi_{\parallel} & = & \arg(A_{\parallel}A^*_0)= - 2.50 \pm 0.20 \pm 0.08,
                 \;\; [{\rm Solution \;II}] \label{phase2}
\end{eqnarray}
is also allowed when $\phi_{\perp,\parallel}$ is chosen in $(-\pi,\pi)$. 
Since $|A_{\parallel}|\approx |A_{\perp}|$ and $\phi_{\parallel}-
\phi_{\perp}\approx \pi$ or $0$, two sets of phases in 
Eqs. (\ref{phase1}) and (\ref{phase2}), referred 
to as Solution I and II, mean roughly 
\begin{equation}
               A_{\parallel}\approx \mp A_{\perp}. \label{exp} 
\end{equation}
That is, either $|H_{+1}|\ll |H_{-1}|$ (Solution I) or 
$|H_{+1}|\gg |H_{-1}|$ (Solution II). To be quantitative, we obtain 
in terms of the helicity amplitudes,
\begin{equation}
  |H_{\pm1}/H_{\mp 1}|=0.26\pm 0.14, \;[{\rm Solution\; I/II}] \label{exp'}
\end{equation}
where the upper and lower signs in the subscripts of the helicity amplitudes
correspond to Solution I and II, respectively.  Our concern is on 
this twofold ambiguity.

\section{Light-quark helicity conservation}

In the decay $B(q\overline{b})\to J/\psi(c\overline{c})K^*(q\overline{s})$  
the $\overline{s}$-quark is produced in helicity $+\frac{1}{2}$ by weak
interaction in the limit of $m_s\to 0$. It would maintain its helicity 
throughout strong interaction if $m_s=0$. Therefore, when the 
$\overline{s}$-quark picks up $q$($u$ or $d$), 
they form $K^*$ in helicity either +1 or 0, not in helicity $-1$. 
Within perturbative QCD this argument is valid as long as we ignore 
corrections of $m_s/E$ and $|{\bf p}_t|/E$, and higher configurations 
of $K^*$ such as $\overline{s}q\overline{q}q$ and $\overline{s}qg$. 
If FSI is entirely of short distances, therefore, the decay amplitudes 
should obey the selection rule;
\begin{equation}
  H_{-1} \simeq 0 \;\;{\rm for}\;B(q\overline{b})\to J/\psi K^*, \label{JK} 
\end{equation}
namely, 
\begin{equation}
 A_{\parallel} \simeq + A_{\perp}\;\;{\rm for}\;B(q\overline{b})
         \to J/\psi K^*. \label{JK'}
\end{equation}
Eq. (\ref{JK'}) means for both magnitude and phase.
Similarly, $H_{+1}\simeq 0$ or $A_{\parallel}\simeq - A_{\perp}$  
for $\overline{B}(\overline{q}b)\to J/\psi\overline{K}^*$. Solution II is 
not far from this prediction. However, validity of the perturbative 
QCD argument is suspect for the decay $B\to J/\psi K^*$ since
the size of $J/\psi$ is $O(1/\alpha_s m_c)$ instead of $O(1/m_c)$\cite{B}. 
If long-distance FSI is important, the $s$-quark helicity can be 
easily flipped through meson-meson rescattering in the final state. 
Then Solution I cannot be ruled out.  

The $B\to J/\psi K^*$ amplitudes were calculated in the past mostly with 
factorization combined with extrapolation or scaling rules of form 
factors\cite{models1,models2,Keum}. Those calculations naturally predicted 
$|H_{+1}|>|H_{-1}|$ for $B\to J/\psi K^*$. Since factorization leads to 
zero strong phases, $|\phi_{\parallel}|-\pi= 37^{\circ}\pm 11^{\circ}
\pm 4^{\circ}$ is a measure of deviation from factorization if Solution I is 
chosen.\footnote{
It was recently pointed out\cite{BH} that the $s$-quark helicity 
conservation is consistent with the decay rate ratio 
$\Gamma(B\to\gamma K^*)/\Gamma(B\to \gamma X_s)$. Without additional
theoretical input, however, experiment on the rates alone cannot 
conclude $h=+1$ dominance.} 

  The case for Solution II may look strong. However, there is no firm
theoretical basis for validity of factorization for $B\to J/\psi K^*$. 
Indeed the observed strong phases are larger than what we would normally 
expect for the short-distance QCD correction to factorization. Furthermore
the Belle Collaboration\cite{Belle} very recently made positive 
identification of the $\overline{B}^0\to D^{(*)0}X^0$ decay modes. 
The branching fraction of $\overline{B}^0\to D^0\pi^0$ is now much larger 
than the tight upper bound that was set by CLEO\cite{Nemati,PDG} and 
advocated by factorization calculation. Those decay modes share one 
common feature with $B\to J/\psi K^*$.  We therefore proceed to explore for 
chance of Solution I, {\em i.e.,} large violation of $s$-quark helicity 
conservation due to large long-distance FSI. 
 
\section{Statistical model of strong phases}

We look for the origin of the fairly large strong phase which is
three standard deviations away from zero.   
One characteristic of the decay $B\to J/\psi K^*$ may be relevant to the 
large phase.  That is, this decay is a color-suppressed process.\footnote{
We mean as usual an $O(1/N_c)$ contribution from the dominant operator
$(\overline{b}c)(\overline{c}s)$ and an $O(1)$ contribution from the 
suppressed operator $(\overline{b}s)(\overline{c}c)$.} A statistical 
model\cite{SW} was proposed for the strong phases of $B$ decay 
for which the short-distance argument fails. The model 
predicts that the more a decay process is suppressed, the larger its strong 
phase can be. The reason is as follows: In a suppressed process of a given 
decay operator, $B$ tends to decay first into unsuppressed decay channels 
and then rescatters into its final state by FSI. In $B\to J/\psi K^*$, 
the $B$ meson decays first into color-allowed on-shell states such as 
$\overline{D}^{(*)}D_s^{(*)}$ and then turns into 
$J/\psi K^*$ through the quark-rearrangement scattering of strong
interactions (crossed quark-line diagram). Such two-step processes are 
likely to dominate over direct color-suppressed transition. If so, 
those on-shell intermediate states tend to generate larger strong 
phases for color-suppressed amplitudes than for color-allowed
amplitudes dominated by the direct transition. 
The same picture was advocated independently by Rosner in his
qualitative argument\cite{Rosner}.

However, computing individual strong phases is a formidable task 
when so many decay channels are open and interact with each other 
through long-distance FSI. The statistical model quantifies 
the range of likely values ($-\overline{\delta}\leq\delta\leq
\overline{\delta}$) for a strong phase $\delta$ in terms of
two parameters, degree of suppression ($1/\rho$) and strengh of FSI 
($\tau$), by the relation\cite{SW}
\begin{equation}
        \tan^2\overline{\delta} = 
        \frac{\tau^2(\rho^2-\tau^2)}{1-\rho^2\tau^2}, \label{SW}
\end{equation}
which is valid for $\tau^2 <\rho^2<1/\tau^2$. Outside this region of $\rho$ 
and $\tau$, the right-hand side of Eq. (\ref{SW}) is negative. In this
case suppression is so severe ($1/\rho^2<\tau^2$) and/or rescattering 
transition between $J/\psi$ and $\overline{D}^{(*)}D_s^{(*)}$ is so strong 
($\tau^2>\rho^2$) that any value is possible for $\delta$.

For the suppression parameter we expect $1/\rho=O(1/N_c)$ in our 
case. Though color suppression does not always work as we expect, 
$1/\rho^2 = O(1/N_c^2)$ is in line with experiment. Let us choose 
$1/\rho^2\simeq 1/20$ by comparing $B(B^+\to J/\psi K^{*+}) = 
(1.48\pm 0.27)\times 10^{-3}$ with $B(B^+\to \overline{D}^{*0}D_s^{*+})
= (2.7\pm 1.0)\times 10^{-2}$\cite{PDG}. To determine the value of $\tau$, we
need strength of $J/\psi K^*$ reaction which is little known. For the total 
cross section, the strength is controlled by Pomeron exchange. Since it is 
generated by two-gluon exchange in the standard lore, one possible 
estimate is $\sigma_{{\rm tot}}^{J/\psi K^*}\approx 
[\alpha_s(E)/\alpha_s(\Lambda_{QCD})]^2\sigma_{{\rm tot}}^{\pi\pi}$ 
where $E=\frac{1}{2}\sqrt{4m_D^2-m^2_{J/\psi}}\simeq$ 
1 GeV is the binding of $J/\psi$. It means that energy transfer of $O(E)$
is needed to break up $J/\psi$ by hitting with a gluon. With this reasoning
we expect rescattering of $J/\psi$ to be less strong than that of $\pi\pi$ 
and $\pi K$. If we choose tentatively $\sigma_{{\rm tot}}^{J/\psi K^*}
\simeq 0.5\times\sigma_{{\rm tot}}^{\pi\pi}$, we find 
$\tau^2\simeq 0.09$\cite{SW}.  For $\rho^2\simeq 20$ and $\tau^2\simeq 0.09$ 
($\rho^2\tau^2\simeq 1.8$), the right-hand side of Eq. (\ref{SW}) 
is negative so that $\delta$ can take any value, as remarked above.
Physically, the cascade processes $B\to\overline{D}^{(*)}D_s^{(*)}
\to J/\psi K^*$ dominate over the direct $B\to J/\psi K^*$ transition
in this case. When this happens, there is no reason to
expect that the $s$-quark helicity conserves. Then it is not impossible 
that $A_{\parallel}$ and $A_{\perp}$ acquire a relative phase large 
enough to flip their relative sign. On the other hand, 
$\sigma_{tot}^{J/\psi K^*}$ may well be much smaller than our estimate 
above. If it is one tenth of $\sigma_{tot}^{\pi\pi}$, for instance, 
the strong phases of $B\to J/\psi K^*$ should be in the range smaller 
than $35^{\circ}$ or so. If this is the case, the direct decay still
dominates and the $s$-quark helicity approximately conserves.

Because of uncertainties in strong interaction physics involved, we are 
unable to make a convincing estimate for likely values of strong phases
of $B\to J/\psi K^*$. We can say only that very large strong phases 
are possible for this decay. We therefore look for other $B$ decay modes 
which will help in resolving the issue.

\section{Spin amplitudes of other $B\to V_1V_2$ modes}

If long-distance FSI is large in $B\to J/\psi K^*$, the pattern of 
\begin{eqnarray}
   |A_{\parallel}| &\simeq& |A_{\perp}| \nonumber \\ 
 \phi_{\parallel}&\simeq& \phi_{\perp}\;({\rm modulo}\; \pi). 
       \label{pattern} 
\end{eqnarray}
must be interpreted as an accident. Measurement of the spin amplitudes for 
$B\to \psi(2s)K^*$ will shed a light in this case: If the same pattern 
appears in $B\to \psi(2s)K^*$, we will favor conservation of $s$-quark 
helicity in the sense that two accidents are rarer to occur than one. 

The decay $B_s\to J/\psi\phi$ is identical to $B\to J/\psi K^*$ up to
$d/u\leftrightarrow s$.  At present we know from CDF\cite{CDF}
\begin{eqnarray}
         |A_0| &=& 0.78 \pm 0.09\pm 0.01 \nonumber \\
         |A_{\parallel}|&=& 0.41\pm 0.23 \pm 0.05 \nonumber \\ 
         |A_{\perp}| &=& 0.48\pm 0.20 \pm 0.04,
\end{eqnarray}
and for the phases
\begin{equation}
      \phi_{\parallel} = \pm 1.1\pm 1.3 \pm 0.2, \label{phase1'} 
\end{equation}
Nothing is known for $\phi_{\perp}$. At present the 
uncertainty of $\phi_{\parallel}$ is too large to make any statement. 
As the experimental uncertainties become smaller, we should watch 
whether $|A_{\parallel}|\approx|A_{\perp}|$ stands or not, 
and whether $\phi_{\parallel}-\phi_{\perp}$ converges to zero 
(modulo $\pi$) or not. If both happen, we can make a stronger case for 
$s$-quark helicity conservation. If either relation is badly violated, 
it will cast doubt on the $s$-quark helicity conservation in 
$B\to J/\psi K^*$. A similar test of the $d$-quark helicity 
conservation in $B\to J/\psi\rho$ will serve for the same purpose. 

The decay mode $B^-\to D^{*0}\rho^-$ provides us an interesting 
opportunity. The decay $\overline{B}^0\to D^{*+}\rho^-$ is a color-allowed 
process ($b\to c_L\overline{u_L}d_L$) for which factorization is expected 
to work well. Here the dominant decay operator is the tree 
operator $(\overline{c}b)(\overline{d}u)$. In this decay $\rho^-$ is 
formed by the collinear $d_L\overline{u_L}$ from the weak current so 
that helicity of $\rho^-$ must be 0, not $\pm 1$. In fact, 
experiment confirmed dominance of $h=0$; $ |A_0|^2/\sum|A_i|^2= 0.93
\pm 0.05 \pm 0.05$\cite{CLEO2}. Since there is only one spin amplitude
of significant magnitude, one cannot measure a strong phase in this mode.
However, validity of perturbative QCD leaves us little doubt about 
the $u/d$-quark helicity conservation and the smallness of the strong phase
in $\overline{B}^0\to D^{*+}\rho^-$.

In contrast, the decay $B^-\to D^{*0}\rho^-$ can occur through a 
color-suppressed process as well since the fast $d_L$ from the weak 
current can pick up the spectator $\overline{u}$ instead of the 
$\overline{u}_L$ from the current. Relative to the dominant process,
this process is not only color-suppressed but also power-suppressed 
through the $\rho^-$ wave function\cite{B}. Despite the expected 
double suppression, this amplitude is not so small in reality 
and shifts square root of the rate by about one third from 
the color-allowed process alone\cite{PDG}:
\begin{equation}
 |\Gamma(B^-\to D^{*0}\rho^-)/\Gamma(\overline{B}^0\to D^{*+}\rho^-)|^{1/2}= 
   1.36 \pm 0.18. \label{exp1}
\end{equation}   
The left-hand side can be expressed as $|1+0.79(a_2/a_1)|$  
in terms of the color-allowed and suppressed amplitudes, 
$a_1$ and $a_2$, in the notation of Bauer, Stech, and Wirbel\cite{BSW}. 
If factorization is a good approximation, $a_{1,2}$ are real and 
$a_2$ is very small ($0< a_2/a_1 < 0.15$) though its precise value 
is sensitive to cancellation between two Wilson coefficients.  
The sizable deviation from unity in the right-hand side of Eq. (\ref{exp1}) 
indicates that the color-suppressed portion of the $B^-\to D^{*0}\rho^-$ 
amplitude exceeds the magnitude predicted by factorization.\footnote{
Although Eq. (\ref{exp1}) alone would allow destructive interference 
between $a_1$ and $a_2$, such a large value for $|a_2|$ would lead us 
to an unacceptably large branching fraction for $\overline{B}^0\to 
D^{*0}\rho^0$ by the $\Delta I = 1$ sum rule.} It can accommodate  
any large phase for $a_2/a_1$. Therefore we should test whether this 
color-suppressed portion of amplitude has a large strong phase or not.

Since $\rho^-$ is dominantly in helicity 0 in the color-allowed
$B^-\to D^{*0}\rho^-$ decay, the helicity amplitudes $H_{\pm 1}$ 
can arise mostly from the color-suppressed decay, if at all. Since $\rho^-$ 
is made of $d_L$ from weak current and the spectator $\overline{u}$ 
in this case, the $\rho^-$ helicity would be either $-1$ or 0, 
not +1. In this respect, the situation is parallel to $B\to J/\psi K^*$ 
up to charge conjugation. The other current quark $\overline{u_L}$
enters $D^{*0}$ so that helicity of $D^{*0}$ must be either +1 or 0
depending on helicity of $c$. Consequently the $u/d$-quark helicity 
conservation would allow only longitudinal meson helicities even in the 
color-suppressed process if short-distance FSI dominates:  
\begin{equation}
    H_{\pm 1} \simeq 0 \;\; {\rm for}\;\; B^-\to D^{*0}\rho^-\;({\rm SD}). 
                               \label{Drho}
\end{equation}
If FSI is entirely of short distances, the expected accuracy of 
Eq. (\ref{Drho}) should be even higher than that of the $s$-quark helicity
conservation. Needless to say that this prediction result in all 
factorization calculations if light-quark helicity conservation is
implemented for form factors. If the pattern of Eq. (\ref{Drho}), namely,
$|A_0|\simeq 1$ emerges in $B^-\to D^{*0}\rho^-$, it will indicate 
short-distance dominance even for its color-suppressed $a_2$ amplitude and 
therefore give an indirect support to the $s$-quark helicity conservation 
in $B\to J/\psi K^*$. For determination of $|A_0|$, we do not need 
full measurement of transversity angular distribution.

Finally we point out that we shall be able to carry out the same test 
with the color-suppressed decay $\overline{B}^0\to D^{*0}\omega$. The 
Belle Collaboration very recently measured this decay branching\cite{Belle} 
at a level much higher than anticipated. We may have a good chance to test  
directly with $\overline{B}^0\to D^{*0}\rho^0$ which consists purely of the
$a_2$ amplitude of $\overline{B}\to D^{*}\rho$. 

\section{Summary} 
 
   We have examined the twofold ambiguity in determination of the spin 
amplitudes of $B\to J/\psi K^*$. One solution is consistent with approximate
$s$-quark helicity conservation despite substantial strong phases, 
while the $s$-quark helicity conservation is badly violated in the other 
solution. Though the case for $s$-quark helicity conservation may look  
stronger to many theorists, a large violation is quite possible at present.
Hence we have explored with the statistical model the possibility of large 
$s$ quark helicity violation and argued how measurement of 
$B\to \psi(2s)K^*$, $B\to J/\psi\phi$, $B^-\to D^{*0}\rho^-$, and 
$\overline{B}^0\to D^{*0}\omega/\rho^0$ will serve toward resolution 
of the issue.

\acknowledgements
I am indebted to H-Y. Cheng, Y-Y. Keum, and S. T'Jampens for important 
communications concerning the sign conventions and the ambiguity in 
determination of the spin amplitudes. I acknowledge conversations with 
G. Burdman and R. N. Cahn. This work was supported in part by Director, 
Office of Science, Office of High Energy and Nuclear Physics, Division of
High Energy Physics of  U.S. Department of Energy under Contract 
DE--AC03--76SF00098 and in part by the National Science Foundation 
under grant PHY--95--14797.

\end{document}